\documentclass[unsortedaddress,nofootinbib,11pt]{revtex4}
\usepackage[utf8]{inputenc}
\usepackage{amsmath,empheq,color}
\usepackage{amsfonts}
\usepackage{amsthm}
\usepackage{amssymb}
\usepackage{graphicx}
\usepackage{hyperref}
\usepackage[normalem]{ulem}
\usepackage{epigraph}
\usepackage{mathrsfs}
\usepackage{bbm}
\usepackage{wrapfig}
\usepackage{lineno}
\usepackage{scalerel,stackengine}

\begin{document}

\title{The symplectic group in Polymer Quantum Mechanics}

\author{Angel Garcia-Chung}
\email{alechung@xanum.uam.mx} 
\affiliation{Departamento de F\'isica, Universidad Aut\'onoma Metropolitana - Iztapalapa\\
San Rafael Atlixco 186, Ciudad de M\'exico 09340, M\'exico}

\begin{abstract}
In this paper we provide the representation of the symplectic group $Sp(2n, \mathbb{R})$ in polymer quantum mechanics. We derive the propagator of the polymer free particle and the polymer harmonic oscillator without considering a polymer scale. The polymer scale is then introduced to reconcile our results with those expressions for the polymer free particle. The propagator for the polymer harmonic oscillator implies non-unitary evolution.
\end{abstract}

\maketitle

\section{Introduction}

Polymer quantum mechanics (PQM) is a quantization scheme mimicking some of the techniques used in Loop Quantum Cosmology (LQC)\cite{ashtekar2003quantum, corichi2007polymer, corichi2007hamiltonian, velhinho2007quantum, pawlowski2014separable}. That is to say, at kinematical level, the Hilbert spaces used in both quantizations are the mathematically the same. Also, their observable algebras are given by their corresponding Weyl algebras, on the reduced phase space for the LQC scenario and on the standard phase space in the PQM case \cite{ashtekar2003quantum, ashtekar2003mathematical, bojowald2010canonical, bojowald2011quantum}. 

The main feature of these quantizations is that the Stone-von Neumann theorem is evaded. Consequently, the quantum description of LQC is non-unitarily equivalent to the so-called Wheeler-De Witt (WDW) quantization \cite{bojowald2011quantum}, whereas PQM is non-unitarily equivalent to the usual Schr\"odinger representation of standard quantum mechanics \cite{ashtekar2003quantum}. Based on this, PQM can be considered as a theoretical lab for some of the techniques used at kinematical level in LQC.

An intrinsic aspect of PQM is the introduction of a length scale, called polymer scale which is the analog of the Planck length in loop quantum cosmology and loop quantum gravity. This scale, denoted by $\mu$, is introduced when the square of the momentum operator in the Hamiltonian is replaced by a combination of Weyl generators depending on $\mu$.

The effect of what this replacement might induce on the symmetries of `polimerically' quantized systems have been explored in some papers \cite{chiou2007galileo, date2013polymer}. However, a key ingredient on this direction is still absent: the linear canonical transformations. These transformations are the symplectic group action of the phase space and leave the Hamilton equations of motion invariant. A relevant feature of this group, denoted by $Sp(2n, \mathbb{R})$, is that time evolution of linear systems can be described as a curve in the group. Therefore, a representation of the symplectic group in PQM will pave the way to study time-evolution and more general linear symmetries.

On the other hand, in the last decades, the symplectic group has played a relevant role in the construction of squeeze states \cite{schnabel2017squeezed, walls2007quantum, adesso2014continuous} and their separability conditions \cite{adesso2007entanglement, braunstein2005quantum}. In cosmological scenarios, some approaches use squeezed states to explore entropy production \cite{gasperini1993quantum} and time-evolution of matter degrees of freedom \cite{grain2019squeezing}. In LQC, Squeezed states have been considered in \cite{ashtekar2006quantum, ashtekar2006quantum2, diener2014numerical} to explore the robustness of the bounce. Due to the connection between the symplectic group and the squeeze operator \cite{adesso2014continuous}, a representation of this group in PQM, and thus in LQC, may offer new insights for constructing generalized squeezed states, i.e., to extend their use to systems with more degrees of freedom.

For these reasons, in this paper we give the representation of the symplectic group $Sp(2n, \mathbb{R})$ in PQM. Our construction was done within the full polymer Hilbert space and no polymer scale was considered at the first stage of the analysis. As a result, the propagators for the free particle and the harmonic oscillator are also derived in the full polymer Hilbert space. We then insert the polymer scale in our results and discuss its implications.

This paper is organized as follows: in Section (\ref{S2}) we provide the unitary representation of the symplectic group in the standard quantum mechanics and its expansion to momentum representation. We also show how the propagators of the quantum free particle and the quantum harmonic oscillator emerge from the given representation. Section (\ref{S3}) summarizes the main features of polymer quantum mechanics to be used for the construction of the representation of $Sp(2n, \mathbb{R})$ given in Section (\ref{S4}). The propagators for both polymer systems, the free particle and the harmonic oscillator are derived in Section (\ref{S5}). We discuss our results in Section (\ref{S6}).

\section{Symplectic group and its standard unitary representation} \label{S2}

The unitary representation of the symplectic group in the Schr\"odinger representation was derived by Mochinsky and Quesne in \cite{moshinsky1971linear} and revisited by K. Wolf in \cite{wolf2016development}. In order to be self-contained, we summarize the construction of the unitary representation of $Sp(2n, \mathbb{R})$ in this section. We follow on the lines of \cite{moshinsky1971linear, wolf2016development} together with \cite{torre2005linear}.

The symplectic group $Sp(2n, \mathbb{R})$ is formed by $2n \times 2n$ matrices ${\bf M}$ of the form
\begin{equation}
{\bf M} = \left(
\begin{array}{cc} 
{\bf A} & {\bf B} \\
{\bf C} & {\bf D}
\end{array}
\right), \label{SGElementM}
\end{equation}
\noindent where ${\bf A}$, ${\bf B}$, ${\bf C}$ and ${\bf D}$ are $n \times n$ matrices such that
\begin{equation}
{\bf A} {\bf D}^T - { \bf B} {\bf C}^T = {\bf 1}, \qquad {\bf A} {\bf B}^T = {\bf B} {\bf A}^T, \qquad {\bf C} {\bf D}^T = {\bf D} {\bf C}^T.
\end{equation}
\noindent These relations result from the symplectic group condition \cite{moshinsky1971linear} given as
\begin{equation}
\left( \begin{array}{cc} {\bf 0} & {\bf 1} \\ -{\bf 1} & {\bf 0} \end{array}\right) = {\bf M} \left( \begin{array}{cc} {\bf 0} & {\bf 1} \\ -{\bf 1} & {\bf 0} \end{array}\right) {\bf M}^{T},
\end{equation}
\noindent where ${\bf M}^{T}$ is the transpose matrix. The group multiplication is the usual matrix multiplication
\begin{equation}
{\bf M}_1 {\bf M}_2 =  \left(
\begin{array}{cc} 
{\bf A}_1 & {\bf B}_1 \\
{\bf C}_1 & {\bf D}_1
\end{array}
\right)  \left(
\begin{array}{cc} 
{\bf A}_2 & {\bf B}_2 \\
{\bf C}_2 & {\bf D}_2
\end{array}
\right) =  \left(
\begin{array}{cc} 
{\bf A}_1 {\bf A}_2 + {\bf B}_1 {\bf C}_2 & {\bf A}_1 {\bf B}_2 + {\bf B}_1 {\bf D}_2 \\
{\bf C}_1 {\bf A}_2 + {\bf D}_1 {\bf C}_2 & {\bf C}_1 {\bf B}_2 + {\bf D}_1 {\bf D}_2
\end{array}
\right),
\end{equation}
\noindent and the inverse of matrix ${\bf M}$, which we denote as ${\bf M}^{-1}$ is given by 
\begin{equation}
{\bf M}^{-1} = \left(
\begin{array}{cc} 
{\bf D}^T & -{\bf B}^T \\
-{\bf C}^T & {\bf A}^T
\end{array}
\right). \label{IMatrix}
\end{equation}
\noindent The identity $2n \times 2n$  matrix is the identity element of this group. 

Some of the important subgroups of this group are: the $SO(n,\mathbb{R})$ group and the non-singular diagonal matrix group $\mbox{diag}(\lambda_1, \lambda_2, \dots, \lambda_n )$ or as we call it the {\it scaling group}.  A rotation ${\bf R} \in SO(n,\mathbb{R})$ can be implemented as an element ${\bf M}(R) \in Sp(2n, \mathbb{R})$ as
\begin{equation}
{\bf M}(R) = \left( \begin{array}{cc} {\bf R} & {\bf 0} \\ {\bf 0} & {\bf R} \end{array}\right),
\end{equation}
\noindent and a scaling matrix ${\bf S} = \mbox{diag}(\lambda_1, \lambda_2, \dots, \lambda_n)$, where $\lambda_j \neq 0$ for $j=1,2,\dots, n$, corresponds to an element ${\bf M}(S) \in Sp(2n, \mathbb{R}) $ given as
\begin{equation}
{\bf M}(S) = \left( \begin{array}{cc} {\bf S} & {\bf 0} \\ {\bf 0} & {\bf S}^{-1} \end{array}\right).
\end{equation}

The standard representation of $Sp(2n, \mathbb{R})$ was derived in \cite{moshinsky1971linear} by Mochinsky and Quesne. They considered the Schr\"odinger representation in the Hilbert space ${\cal H}^{(q)} = L^2(\mathbb{R}^n, d \vec{x})$ given by the fundamental operators
\begin{equation}
\vec{\widehat{q}} \, \Psi(\vec{x}) = \vec{x} \, \Psi(\vec{x}), \qquad \vec{\widehat{p}} \, \Psi(\vec{x}) = \frac{\hbar}{i} \vec{\nabla}^{(q)} \Psi(\vec{x}), \label{SchRepCoord} 
\end{equation}
\noindent where $\vec{\widehat{q}}$  stands for a horizontal array of operators $(\widehat{q}_1, \widehat{q}_2, \dots, \widehat{q}_n)$ and similarly, $\vec{\widehat{p}}$ and $\vec{\nabla}^{(q)}$ stand for $(\widehat{p}_1, \widehat{p}_2, \dots, \widehat{p}_n)$ and $(\partial_{x_1}, \partial_{x_2}, \dots, \partial_{x_n})$ respectively.

The condition for the group action in ${\cal H}^{(q)}$ is of the form
\begin{equation}
\widehat{C}_M  \left(
\begin{array}{c} 
\vec{\widehat{q}}^T   \\
\vec{\widehat{p}}^T
\end{array}
\right) \widehat{C}^{-1}_M = {\bf M}^{-1} \left(
\begin{array}{c} 
\vec{\widehat{q}}^T   \\
\vec{\widehat{p}}^T
\end{array} 
\right), \label{MochisnkyCondition}
\end{equation}
\noindent where $\widehat{C}_M$ is the quantum operator associated to the group element ${\bf M}$.  In (\ref{MochisnkyCondition}) the operator products on the left are taken component by component in the $\vec{\widehat{q}}$ and $\vec{\widehat{p}}$ and the matrix ${\bf M}^{-1}$ on the right is given in (\ref{IMatrix}). The action of the group element $\widehat{C}_M$ on the state $\Psi(\vec{x})$ is
\begin{equation}
\Psi_M(\vec{x}) = \widehat{C}_M \Psi(\vec{x}) = \int d^n \vec{x}' C_{M}(\vec{x}, \vec{x}') \Psi(\vec{x}').
\end{equation}
\noindent Here, $\Psi_M(\vec{x})$ is the new state in $L^2(\mathbb{R}^n, d\vec{x})$ and the kernel $C_{M}(\vec{x}, \vec{x}') $ satisfies the relation
\begin{equation}
C_{M_ 1 \, M_2}(\vec{x}, \vec{x}'') = \int d^n \vec{x}' C_{M_1}(\vec{x}, \vec{x}') C_{M_2}(\vec{x}', \vec{x}''),
\end{equation}
\noindent and is such that the operator $\widehat{C}_M$ is unitary. The expression for this kernel, as showed by K. Wolf in \cite{wolf2016development} is
\begin{equation}
C_M(\vec{x}, \vec{x}') = \frac{1}{ \sqrt{  (2 \pi i \hbar )^n \det {\bf B} }} e^{ \frac{i}{2 \hbar}  \left[ \vec{x}^T {\bf D}{\bf B}^{-1} \vec{x} - 2 \vec{x}'^T {\bf B}^{-1} \vec{x}  + \vec{x}'^T{\bf B}^{-1} {\bf A} \vec{x}'    \right] }, \label{CRMochinsky}
\end{equation}
\noindent and recall that ${\bf A}$, ${\bf B}$ and ${\bf D}$ are given by ${ \bf M}$ via (\ref{SGElementM}).

In the present work, the analysis for the group representation requires both representations, the coordinate representation and the momentum representation in polymer quantum mechanics. For this reason and in order to be self-contained, we show some of the main ingredients of the momentum representation of the symplectic group in the standard quantum mechanics. To do so, recall that the Hilbert space in the momentum representation is ${\cal H}^{(p)} = L^2(\mathbb{R}^n, d\vec{p})$ and the fundamental operators are given as
\begin{eqnarray}
&& \vec{\widehat{q}} \, \widetilde{\Psi}(\vec{p}) = i \hbar  \vec{\nabla}^{(p)} \widetilde{\Psi}(\vec{p}), \qquad \vec{\widehat{p}} \, \widetilde{\Psi}(\vec{p}) = \vec{p} \, \widetilde{\Psi}(\vec{p}),  \label{SchRepMomen}
\end{eqnarray}
\noindent where $\vec{\widehat{p}}$\,  is now a horizontal array of operators $(\widehat{p}_1, \widehat{p}_2, \dots, \widehat{p}_n)$ whereas $\vec{\widehat{q}}$ and $\vec{\nabla}^{(p)}$ are $(\widehat{q}_1, \widehat{q}_2, \dots, \widehat{q}_n)$ and $(\partial_{p_1}, \partial_{p_2}, \dots, \partial_{p_n})$ respectively.

To derive the expression for the kernel we can proceed in a twofold manner: (i) we can perform the Fourier transform of the kernel in (\ref{CRMochinsky}) or (ii) we can reproduced the steps given in \cite{moshinsky1971linear, wolf2016development} but in the momentum representation. The result is the same and once the calculation is carried out, we obtain the expression for the new kernel in ${\cal H}^{(p)}$ to be of the form
\begin{equation}
\widetilde{C}_M(\vec{p}, \vec{p}') = \sqrt{ \left( \frac{i}{2 \pi \hbar} \right)^n  \frac{1}{\det C} } \; e^{   - \frac{i}{2 \hbar}  \left[  \vec{p}^T {\bf A}{\bf C}^{-1} \vec{p} - 2 \vec{p}'^T {\bf C}^{-1} \vec{p} + \vec{p}'^T {\bf C}^{-1} {\bf D}\vec{p}'    \right] }. \label{MRMochinsky}
\end{equation}

On both equations (\ref{CRMochinsky}) and (\ref{MRMochinsky}) singular matrices ${\bf B}$ and ${\bf C}$ produce a singular value of the kernel. To overcome this issue, Mochinksky and Quesne \cite{moshinsky1971linear} proved that every matrix ${\bf M}$ with a singular matrix ${\bf B}$ can be written as the product $ {\bf M} = {\bf M}' \; \widetilde{\bf M}  $, where the matrices ${\bf M}' $ and $ \widetilde{\bf M} $, given as
\begin{equation}
{\bf M}' = \left( \begin{array}{cc} {\bf 1} & {\bf B}' \\ {\bf 0} & {\bf 1} \end{array}\right), \quad \widetilde{\bf M}= \left( \begin{array}{cc} {\bf A} - {\bf B}' {\bf C} & {\bf B} - {\bf B}' {\bf D} \\ {\bf C} & {\bf D} \end{array}\right) ,
\end{equation}
\noindent are such that the matrix ${\bf B}'$ is non-singular and diagonal and the matrix $ {\bf B} - {\bf B}' {\bf D} $ is non-singular. Therefore, the kernel in (\ref{CRMochinsky}) is well defined for these matrices. A similar analysis can be carried out for the kernel in (\ref{MRMochinsky}) but we will omit it in this work since it is not required.

\subsection{Propagator analysis}

In this subsection we derive the propagator for two mechanical systems using the given representation of $Sp(2n, \mathbb{R})$. Both examples are linear Hamiltonian systems since their time evolution is a matrix belonging to the symplectic group. The first example is the $n=1$ free particle of mass $m$. The solution of the classical Hamilton equations for this system are
$$ q(t) = q + \frac{t}{m} p, \qquad p (t) = p,$$
\noindent and can be written in matrix form as follows
\begin{equation}
\left( \begin{array}{c} q(t)  \\ p(t)  \end{array}\right) = {\bf M}_{FP}(t) \left( \begin{array}{c} q  \\ p  \end{array}\right),
\end{equation}
\noindent where the matrix ${\bf M}_{FP}(t)$ takes the form
\begin{equation}
{\bf M}_{FP}(t) = \left( \begin{array}{cc} 1 & \frac{t}{m} \\ 0 & 1 \end{array} \right). \label{PLP}
\end{equation}

Notice that ${\bf M}_{FP}(t)$ is not just a matrix but also a curve in $Sp(2,\mathbb{R})$, i.e., it is a map $t : \mathbb{R} \rightarrow Sp(2,\mathbb{R})$. From (\ref{PLP}) we read $A=D=1$, $C=0$ and $B= t/m$. After inserting these expressions in (\ref{CRMochinsky}) together with $n=1$ we have
\begin{equation}
C_{FP}(x,x'; t) = \sqrt{ \frac{m}{2 \pi i \hbar t} } e^{ \frac{i \, m}{2 \hbar t} (x-x')^2}, \label{FPPropagator}
\end{equation}
\noindent which is the propagator of the quantum free particle. Recall that no Hamiltonian operator $\widehat{H}$ was used to derive the propagator in (\ref{FPPropagator}),  although it is defined as the kernel $\langle x | e^{\frac{i}{\hbar} \Delta t \widehat{H}} | x' \rangle $. This is the main idea we want to exploit using the representation of the symplectic group in polymer quantum mechanics.

The second example is the $n=1$ harmonic oscillator for which the Hamilton equations
$$ q(t) = q \cos(\omega \, t) + \frac{p}{m \, \omega} \sin(\omega \, t), \qquad 
p(t) = - m \omega q \sin(\omega \, t) + p \cos(\omega \, t), $$
\noindent where $m$ and $\omega$ are respectively, the mass and the oscillator frequency. When these equations are written in matrix form they gives the matrix ${\bf M}_{HO}(t)$ as
\begin{equation}
{\bf M}_{HO}(t) = \left( \begin{array}{cc} \cos(\omega t) & \frac{1}{m \omega} \sin(\omega t ) \\ - m \omega \sin(\omega t) & \cos(\omega t)  \end{array}\right). \label{SMHO}
\end{equation}
\noindent  Plugging in (\ref{CRMochinsky}) the expressions for $A=D=\cos(\omega t)$ and $B= \frac{1}{m \omega} \sin(\omega t)$ and $C = - m \omega \sin(\omega t) $ we get
\begin{equation}
C_{HO}(x,x'; t) = \sqrt{ \frac{m \omega}{ 2 \pi i \hbar \sin(\omega t)}  } e^{ \frac{i \, m  \omega}{2 \hbar \sin(\omega t)}   \left[ (x^2 + x'^2 ) \cos(\omega t) - 2 x \, x'\right] }, \label{QHOPropagator}
\end{equation}
\noindent which is the propagator for the quantum harmonic oscillator.

The expression for the propagator in the momentum representation can also be derived using the formula in (\ref{MRMochinsky}). In the case of the quantum harmonic oscillator it takes the form
\begin{equation}
\widetilde{C}_{HO}(p, p'; t) = \sqrt{ \frac{1}{ 2 \pi i \hbar m \omega \sin(\omega t)}  } e^{ \frac{i}{2 \hbar m \omega \sin(\omega t)}   \left[ (p^2 + p'^2 ) \cos(\omega t) - 2 p \, p' \right] }. \label{PHOPropagator}
\end{equation}

The expressions for the propagators (\ref{FPPropagator}) and (\ref{QHOPropagator}), together with (\ref{PHOPropagator}) were derived using the classical equations of motion. The quantum dynamics has not yet been considered, only the Schr\"odinger representations (\ref{SchRepCoord}) and (\ref{SchRepMomen}) were considered to obtain these propagators. 

The aim of this subsection was to show the advantage of having the (unitary) representation of the symplectic group: it allows to derive the quantum propagator for linear Hamiltonian systems. Of course, all the analysis was done without explicitly knowing the representation of the Hamiltonian operator. In the next sections, we will implement this analysis in the polymer quantum mechanics and then we will derive the polymer propagators for the free particle and the quantum harmonic oscillator.

\section{Representations in Polymer Quantum Mechanics} \label{S3}

In polymer quantum mechanics \cite{ashtekar2003quantum, corichi2007polymer, corichi2007hamiltonian, velhinho2007quantum, pawlowski2014separable}, ${\cal H}^{(q)}_{poly} = L^2(\mathbb{R}_d, dx_c) $ is the Hilbert space used in the coordinate representation. Here, $\mathbb{R}_d$ is the real line with discrete topology whereas $dx_c$ is the countable measure on it. An arbitrary state in this Hilbert space can be written as
\begin{equation}
\Psi(x) = \sum_{\{ x_j \}} \Psi_{x_j} \delta_{x, x_j} , \label{Pstate}
\end{equation}
\noindent i.e., a complex-valued function on $\mathbb{R}_d$ such that (i) the coefficients $\Psi_{x_j}$ vanish at all but a countable number of points and (ii) they can be used to define the norm of the state as
\begin{equation}
\sum_{\{ x_j \} } | \Psi_{x_j} |^2 < \infty. \label{Norm}
\end{equation}

 In this notation, $\{ x_j \}$ is an arbitrary graph of numbers in the real line and the coefficients $\Psi_{x_j} \in \mathbb{C}$ are non-zero only on a countable subset denoted as $\{ \tilde{x}_j \}^{\infty}_{j=1}$. Also, the basis in ${\cal H}^{(q)}_{poly}$ is uncountable which can be seen from the state (\ref{Pstate}) where the Kronecker deltas $\delta_{x, x_j}$ are the basis elements. 

This construction can be expanded for a system with $n$ degrees of freedom and a Hilbert space given by $ {\cal H}^{(q)}_{(n) poly} = L^2(\mathbb{R}^n_d, d\vec{x}_c) $. Again, an arbitrary state is of the form
\begin{equation}
\Psi(\vec{x}) = \sum_{\{ \vec{x}_j \}} \Psi_{\vec{x}_j} \delta_{\vec{x}, \vec{x}_j},
\end{equation}
\noindent where, similarly to the $n=1$ case, $\{\vec{x}_j\}$ represents a graph in the $\mathbb{R}^n_d$ space with discrete topology and $d\vec{x}_c$ is the countable measure on it.

The fundamental operators are described using the Weyl algebra elements $\widehat{W}(\vec{a}, \vec{b})$ which, for a system with $n$ degrees of freedom satisfies the canonical commutation relations
\begin{equation}
\widehat{W}(\vec{a}_1, \vec{b}_1) \widehat{W}(\vec{a}_2, \vec{b}_2) = e^{- \frac{i}{2 \hbar} \left( \vec{a}^T_1 \vec{b}_2 - \vec{b}^T_1 \vec{a}_2 \right)} \widehat{W}(\vec{a}_1 + \vec{a}_2, \vec{b}_1 + \vec{b}_2).
\end{equation}
\noindent The real arrays $\vec{a} = (a_1, a_2, \dots, a_n)$ and $\vec{b} = (b_1, b_2, \dots, b_n)$, which have dimensions $[a_j] = \mbox{momentum}$ and $[b_j] = \mbox{coordinate}$, label the Weyl algebra generators. The representation of the generators in ${\cal H}^{(q)}_{(n) poly}$ is  
\begin{equation}
\widehat{W}(\vec{a}, \vec{b}) \Psi(\vec{x}) = e^{ \frac{i}{2 \hbar} \vec{a}^T \vec{b}} e^{ \frac{i}{\hbar} \vec{a}^T \vec{x}} \Psi(\vec{x} + \vec{b}), \label{PCRep}
\end{equation}
\noindent and notice the parameters $\vec{b}$ are related with finite translations. For this reason, the Weyl algebra generators $\widehat{W}(\vec{a}, \vec{b})$ are split in two types of generators
\begin{equation}
\widehat{U}(\vec{a}) := \widehat{W}(\vec{a}, 0), \quad \widehat{V}(\vec{b}) := \widehat{W}(0, \vec{b}).
\end{equation}
\noindent The operator $\widehat{V}$ is responsible for finite translations in the coordinate representation as can be seen from (\ref{PCRep}) by taking $\vec{a} = 0$
\begin{equation}
\widehat{V}(\vec{b}) \Psi(\vec{x}) = \Psi(\vec{x} + \vec{b}).
\end{equation}
\noindent Becaause the representation is not weakly continuous, there are no infinitesimal translations operator, i.e., there are no $\widehat{p}_j$ operators. To circumvent this issue and in order to set a dynamical description, $\widehat{V}(\vec{b})$ is used to provide a replacement for the square momentum operator. To do so, a polymer scale is considered, i.e., it is assumed the existence of a `fundamental' length scale, denoted as $\mu$ which mimics the polymer scale in loop quantum cosmology but, in this case, the value of this fundamental scale is rather arbitrary although it has an upper bound.  This upper bound on $\mu$ is for consistency with basic observations  \cite{ashtekar2003quantum}.

 Once this polymer scale $\mu$ is considered, the replacement for the square momentum is given as the following combinations of $\widehat{V}(\vec{\mu})$ operators
\begin{equation}
\widehat{p^2}_{\mu} := \frac{\hbar^2}{\mu^2} \left[ 2 - \widehat{V}(\vec{\mu}) - \widehat{V}^\dagger(\vec{\mu})  \right], \label{SPProposal}
\end{equation}
\noindent where $\vec{\mu} := (\mu, \mu, \dots, \mu)$. Of course, some might consider a polymer scale for each of the degrees of freedom but in this work we restrict ourselves to consider only one for the sake of the simplicity in the construction.

Naturally, there are plenty of combinations of the $\widehat{V}(\vec{\mu})$, for example, 
\begin{equation}
\widehat{p^2}_{\mu} := \frac{\hbar^2}{12 \mu^2} \left[ 30 + \widehat{V}(2 \vec{\mu}) + \widehat{V}(-2\vec{\mu}) - 16 \widehat{V}(\vec{\mu}) - 16 \widehat{V}(-\vec{\mu})  \right], \label{EPProposal}
\end{equation}
\noindent which will render a dynamical description completely different to that given by the proposal (\ref{SPProposal}). This is what is considered as the Hamiltonian ambiguity in polymer quantum mechanics.

The usual approach to this issue is to consider that a well defined $\widehat{p^2}_\mu$ proposal should correspond, {\it in the `limit' when the polymer scale is far smaller than the intrinsic length of the physical system}, to the quantum description of the standard quantum mechanics. Nevertheless, despite this limit-criterion, (polymer) quantum systems described with Hamiltonian (\ref{SPProposal}) and (\ref{EPProposal}) are going to be entirely different: polymer free particle with (\ref{SPProposal}) is not the same physical system as a polymer free particle using (\ref{EPProposal}) and yet, we consider both as `polymer free particle'. Moreover, the symmetries can be drastically altered when different Hamiltonians are in use.

In the next section we will come back to this analysis and point out some of the advantages of the representation of the symplectic group in polymer quantum mechanics.

\subsection{Momentum representation}

Similarly to the standard quantum mechanics description, we will briefly describe the momentum representation in polymer quantum mechanics. The Hilbert space used for the momentum representation is given by
\begin{equation}
{\cal H}^{(p)}_{poly} = L^2(\overline{\mathbb{R}}, dx_{Bohr}),
\end{equation}
\noindent where $\overline{\mathbb{R}}$ is the Bohr compactification of the real line and $dx_{Bohr}$ is the Bohr measure on this space (see \cite{velhinho2007quantum} for more details). As before, an arbitrary state can be written as
\begin{equation}
\widetilde{\Psi}(p) = \sum_{\{ x_j \}} \Psi_{x_j} e^{ \frac{i}{\hbar} x_j p },
\end{equation}
\noindent and the set $\{ x_j \}$ and the coefficients $\Psi_{x_j}$ satisfy the same conditions mentioned in the analysis of the coordinate representation. The norm in this space is
\begin{equation}
|| \tilde{\Psi} ||^2 = \lim_{L \rightarrow \infty} \frac{1}{2L} \int^L_{-L} |\tilde{\Psi}(p)|^2 dp ,
\end{equation}
\noindent and coincides with (\ref{Norm}). 

The Weyl algebra representation is
\begin{equation}
\widehat{W}(\vec{a}, \vec{b}) \widetilde{\Psi}(\vec{p}) = e^{  - \frac{i}{2 \hbar} \vec{a}^T \vec{b}} e^{- \frac{i}{\hbar} \vec{b}^T \vec{p}} \widetilde{\Psi}(\vec{p} + \vec{a}),
\end{equation}
\noindent and notice that now, the translation is induced with the parameter $\vec{a}$ which implies the operator $\widehat{U}(\vec{a})$ is giving rise to infinitesimal translations but in momentum space. The operator $\widehat{V}(\vec{b})$ acts multiplicatively in this representation and, again, due to it is non-weakly continuous, there is no momentum operator $\widehat{p}$.

For systems with $n$ degrees of freedom, the Hilbert space used for the momentum representation is given as
\begin{equation}
{\cal H}^{(p)}_{(n) poly} = L^2(\mathbb{\overline{R}}^n, d\vec{p}_{Bohr}),
\end{equation}
\noindent and a general element in this space is
\begin{equation}
\tilde{\Psi}(\vec{p}) = \sum_{ \{ \vec{x}_j \}} \Psi_{\vec{x}_j} e^{ \frac{i}{\hbar} \vec{p}^T \vec{x}_j}.
\end{equation}
The physics on both representations is the same, and this can be stated by introducing what is call a Fourier transform ${\cal F}$ between these representations
\begin{eqnarray}
\widetilde{\Psi}(p) &=& {\cal F}[\Psi(x)] =\sum_x e^{\frac{i}{\hbar} x p } \Psi(x), \\
\Psi(x) &=& {\cal F}^{-1}[\widetilde{\Psi}(p)]= \lim_{L \rightarrow \infty} \frac{1}{2L} \int^L_{-L} \widetilde{\Psi}(p) e^{ - \frac{i}{\hbar} x p} dp,
\end{eqnarray}
\noindent for $\Psi(x) \in {\cal H}^{(q)}_{poly}$ and $\widetilde{\Psi}(p) \in {\cal H}^{(p)}_{poly}$. This Fourier transformation can be extended to $n-$degrees of freedom using the replacement
\begin{equation}
\sum_x e^{\frac{i}{\hbar} x \,p} \quad \Rightarrow \quad \sum_{\vec{x}} e^{\frac{i}{\hbar} \vec{x}^T \,\vec{p}},
\end{equation}
\noindent and its inverse can be similarly defined.

\subsection{Distribution analysis}

In the next section we use an ansatz to derive the representation of the symplectic group. To clarify our proposal for this ansatz let us show some features needed for its derivation.

Consider ${\cal H}^{(q)}_{poly}$ and the arbitrary state (\ref{Pstate}). The identity operator $\widehat{1}$ acting on this Hilbert space is such that $ \widehat{1} \Psi(x) = \Psi(x)$, this can also be written as an integral operator 
\begin{equation}
\Psi(x) = \widehat{1} \Psi(x) = \sum_{x'} \delta_{x,x'} \Psi(x') = \int_{\mathbb{R}_d} \delta_{x,x'} \Psi(x') dx_c, \label{UnityOp}
\end{equation}
\noindent for any $\Psi(x) \in {\cal H}^{(q)}_{poly}$. The Kronecker delta $\delta_{x,x'}$ can be consider as a distribution in (\ref{UnityOp}). Our goal is then to find the Fourier transform of the kernel $\delta_{x,x'}$ in the momentum representation. The Fourier transform of $\delta_{x,x'}$ is
\begin{equation}
\lambda(p,p') = \sum_{x, x'} e^{\frac{i}{\hbar} x\, p } \delta_{x,x'} e^{-\frac{i}{\hbar} x' \, p' } = \sum_x e^{\frac{i}{\hbar} x\, (p-p') }, \label{KIOp}
\end{equation}
\noindent where the summation $\sum_{x}$ naturally appears once the Kronecker delta is evaluated.

 It can be checked that indeed
\begin{equation}
\widetilde{\Psi}(p) = \widehat{1} \widetilde{\Psi}(p) = \lim_{L \rightarrow \infty} \frac{1}{2L} \int^{L}_{-L} \lambda(p,p') \widetilde{\Psi}(p') dp', \label{LambdaMomentum}
\end{equation}
\noindent hence, $\lambda(p,p')$ is the kernel of the identity operator $\widehat{1}$ in the momentum representation. Notice the uncountable summation $\sum_x$ in (\ref{KIOp}) is a result of the Fourier transform on each of the arguments of the Kronecker delta and that it is well defined as (\ref{LambdaMomentum}) shows. Moreover, this expression can be expanded to a system with $n$-degrees of freedom as
\begin{equation}
\lambda(\vec{p}, \vec{p}')  = \sum_{\vec{x}} e^{\frac{i}{\hbar} \vec{p}^T \vec{x} - \frac{i}{\hbar} \vec{p}'^T \vec{x} }. \label{Indentity}
\end{equation}

 This summation-like structure in (\ref{Indentity}), when restricted to integer numbers is known as `Dirac comb'. However, in the present case, the summation index $\vec{x}$ runs over the real numbers \cite{prieto2013band, fewster2008phase}. When written in the integral form, the summation in (\ref{Indentity}) for $n=1$ takes the form
\begin{equation}
\sum_{x} e^{\frac{i}{\hbar} (p  - p' ) x} = \int_{\mathbb{R}_d} e^{\frac{i}{\hbar} (p  - p' ) x} dx_c, \label{IntFormalNot}
\end{equation}
\noindent where $\mathbb{R}_d$ stands for the discrete topology of the real line and $dx_c$ is the counting measure. What we want to emphasize is that this measure is scale invariant: if we take $x=a \tilde{x}$, where $a \neq 0$ is a real number, then we have 
\begin{equation}
\sum_{x} e^{\frac{i}{\hbar} (p  - p' ) x} = \sum_{\tilde{x}} e^{\frac{i}{\hbar} (p  - p' ) a \tilde{x}} = \sum_{x} e^{\frac{i}{\hbar} (p  - p' ) a x} ,
\end{equation}
\noindent and if we use the integral formal notation in (\ref{IntFormalNot}) this gives
\begin{equation}
\int_{\mathbb{R}_d} e^{\frac{i}{\hbar} (p  - p' ) x} dx_c = \int_{\mathbb{R}_d} e^{\frac{i}{\hbar} (p  - p' ) a \tilde{x}} d\tilde{x}'_c,
\end{equation}
\noindent which implies $\int_{\mathbb{R}_d} dx_c = \int_{\mathbb{R}_d} d \tilde{x}'_c$ for $x$ and $x'$ scale-related. 

Finally, we can conclude that distributions in the polymer Hilbert space $L^2(\overline{\mathbb{R}}, dp_{Bohr})$ can have this uncountable summation structure, as we will see in the case of the propagators of the free particle and the harmonic oscillator in the last section.

\section{Polymer representation of the symplectic group} \label{S4}

The first step to give the representation of the symplectic group in polymer quantum mechanics is to re-write the condition (\ref{MochisnkyCondition}) but in terms of the Weyl algebra generators $\widehat{W}(\vec{a}, \vec{b})$. By exponentiating each condition in (\ref{MochisnkyCondition}) we have that
\begin{eqnarray}
\widehat{C}^{(poly)}_M \widehat{W}(\vec{a}, 0) \widehat{C}^{(poly) -1}_M &=& \widehat{W}({\bf D} \vec{a}, - {\bf B} \vec{a}),  \label{CondPoly1}\\
\widehat{C}^{(poly)}_M \widehat{W}(0, \vec{b}) \widehat{C}^{(poly) -1}_M &=& \widehat{W}(- {\bf C} \vec{b}, {\bf A} \vec{b}). \label{CondPoly2}
\end{eqnarray}

Analogously to $\widehat{C}_M$ in the standard quantum mechanics, the operator $\widehat{C}^{(poly)}_M$ is the representation of the group element $M$ in polymer quantum mechanics and its action is given by
\begin{equation}
\Psi_M(\vec{p}) := (\widehat{C}^{(poly)}_M \Psi)(\vec{p}) = \lim_{L \rightarrow \infty} \frac{1}{2L} \int^L_{-L} d^n \vec{p}' \, C^{(poly)}_M(\vec{p}, \vec{p}') \Psi(\vec{p}),
\end{equation}
\noindent and $C^{(poly)}_M(\vec{p}, \vec{p}')$ is the unknown polymer kernel.
 
We now multiply on both sides of equation (\ref{CondPoly1}) by the state $\Psi_M(\vec{p})$. On the LHS we have
\begin{equation}
\widehat{C}^{(poly)}_M \widehat{W}(\vec{a}, 0) \widehat{C}^{(poly) -1} _M \Psi_M(\vec{p}) = \widehat{C}^{(poly)}_M   \Psi(\vec{p} + \vec{a}) = \lim_{L \rightarrow \infty} \frac{1}{2L} \int^L_{-L} d^n \vec{p}'\, C^{(poly)}_M(\vec{p}, \vec{p}') \Psi(\vec{p}' + \vec{a}) , \label{LHS1}
\end{equation}
\noindent and on the RHS the expression
\begin{equation}
\widehat{W}({\bf D} \vec{a}, - {\bf B} \vec{a}) \Psi_M(\vec{p}) = e^{\frac{i}{2 \hbar}\vec{a}^T {\bf D}^T {\bf B} \vec{a}  }  e^{ \frac{i}{\hbar} \vec{a}^T {\bf B}^T \vec{p}}  \lim_{L \rightarrow \infty} \frac{1}{2L} \int^L_{-L} d^n \vec{p}' \, C^{(poly)}_M(\vec{p} + {\bf D}\vec{a}, \vec{p}' ) \Psi(\vec{p}') , \label{RHS1}
\end{equation}
\noindent hence equating (\ref{LHS1}) with (\ref{RHS1}) gives the resulting condition to be of the form
\begin{eqnarray}
\lim_{L \rightarrow \infty} \frac{1}{2L} \int^L_{-L} d^n \vec{p}'\, C^{(poly)}_{M}(\vec{p}, \vec{p}') \widetilde{\Psi}(\vec{p}' + \vec{a})  = e^{ \frac{i}{2 \hbar} \vec{a}^T {\bf D}^T {\bf B}\vec{a} } e^{\frac{i}{\hbar} \vec{a}^T {\bf B}^T \vec{p} } \lim_{L \rightarrow \infty} \frac{1}{2L} \int^L_{-L} d^n \vec{p}' \, C^{(poly)}_M(\vec{p} + {\bf D} \vec{a}, \vec{p}') \Psi(\vec{p}')  . \nonumber  \\
\label{Cond1}
\end{eqnarray}

Following the same steps with the expression (\ref{CondPoly2}) we obtain the condition
\begin{eqnarray}
\lim_{L \rightarrow \infty} \frac{1}{2L} \int^L_{-L} d^n \vec{p}'\,  C^{(poly)}_{M}(\vec{p}, \vec{p}') e^{ - \frac{i}{\hbar}\vec{b}^T \vec{p}' }  \widetilde{\Psi}(\vec{p}')  = e^{ \frac{i}{2 \hbar} \vec{b}^T {\bf C}^T {\bf A}\vec{b} } e^{- \frac{i}{\hbar} \vec{b}^T {\bf A}^T \vec{p} } \lim_{L \rightarrow \infty} \frac{1}{2L} \int^L_{-L} d^n \vec{p}' \, C^{(poly)}_M(\vec{p} - {\bf C} \vec{b}, \vec{p}') \Psi(\vec{p}')  . \nonumber \\
\label{Cond2}
\end{eqnarray}

Let us now consider the ansatz
\begin{equation}
C^{(poly)}_M(\vec{p}, \vec{p}') = k^{(poly)}_M e^{ -\frac{i}{2 \hbar} \vec{p}^T  \alpha \vec{p}} \sum_{\vec{x}} e^{ \frac{i}{\hbar} \vec{p}^T \vec{x}} e^{ - \frac{i}{\hbar} \vec{p}'^T \gamma \vec{x} } e^{ \frac{i}{2\hbar} \vec{x}^T \delta \vec{x}}, \label{ansatz}
\end{equation}
\noindent where $\alpha$, $\gamma$ and $\delta$ are $n\times n$ matrices to be determined and $k^{(poly)}_M$ is an unknown complex-valued factor. After inserting this ansatz (\ref{ansatz}) in the condition (\ref{Cond1}) we obtain that
\begin{equation}
\alpha = {\bf B} {\bf D}^{-1}, \qquad \gamma = {\bf D}^T.
\end{equation}
With these results for $\alpha$ and $\gamma$ the kernel now reads as
\begin{equation}
C^{(poly)}_M(\vec{p}, \vec{p}') = k^{(poly)}_M e^{ -\frac{i}{2 \hbar} \vec{p}^T  {\bf B} {\bf D}^{-1} \vec{p}} \sum_{\vec{x}} e^{ \frac{i}{\hbar} \vec{p}^T \vec{x} - \frac{i}{\hbar} \vec{p}'^T {\bf D}^t \vec{x} + \frac{i}{2\hbar} \vec{x}^T \delta \vec{x}} . \label{TempKernel}
\end{equation}
Proceeding now with the second condition (\ref{Cond2}) we insert (\ref{TempKernel}) and obtain the expression for the matrix $\delta$ as
\begin{equation}
\delta = { \bf D} {\bf C}^T,
\end{equation}
\noindent which yields the following expression for the kernel
\begin{equation}
C^{(poly)}_M(\vec{p}, \vec{p}') = k^{(poly)}_M e^{ -\frac{i}{2 \hbar} \vec{p}^T  {\bf B} {\bf D}^{-1} \vec{p}} \sum_{\vec{x}} e^{ \frac{i}{\hbar} \vec{p}^T \vec{x} - \frac{i}{\hbar} \vec{p}'^T {\bf D}^T \vec{x} + \frac{i}{2\hbar} \vec{x}^T {  { \bf D} {\bf C}^T  } \vec{x}}. \label{AnsatzKM}
\end{equation}

To determine the coefficient $k^{(poly)}_M$ we calculate first the kernel of the operator $\widehat{C}^{(poly)}_{M_3}$, where $M_3 = M_1 \cdot M_2$. This allows us to relate this kernel $C^{(poly)}_{M_3}$ with the kernel of the operators $\widehat{C}^{(poly)}_{M_1}$ and $\widehat{C}^{(poly)}_{M_2}$ such that $ \widehat{C}^{(poly)}_{M_3} = \widehat{C}^{(poly)}_{M_1} \cdot \widehat{C}^{(poly)}_{M_2}$. The relation is given by the expression
\begin{eqnarray}
C^{(poly)}_{M_3}(\vec{p}, \vec{p}') = \lim_{L \rightarrow \infty} \frac{1}{2L} \int^L_{-L} d^n \vec{p}'' C^{(poly)}_{M_1}(\vec{p}, \vec{p}'') C^{(poly)}_{M_2}(\vec{p}'', \vec{p}') ,
\end{eqnarray}
\noindent and gives the following condition for the kernels multiplication
\begin{eqnarray}
C^{(poly)}_{M_3 = M_1 M_2 }(\vec{p}, \vec{p}') &=& k^{(poly)}_{M_1 M_2} e^{ -\frac{i}{2 \hbar} \vec{p}^T  ({\bf A}_1 {\bf B}_2 + {\bf B}_1 {\bf D}_2 )({\bf C}_1 {\bf B}_2 + {\bf D}_1 {\bf D}_2)^{-1} \vec{p}} \sum_{\vec{x}} exp\left[ { \frac{i}{\hbar} \vec{p}^T \vec{x}  - \frac{i}{\hbar} \vec{p}'^T ({\bf C}_1{\bf B}_2 + {\bf D}_1 {\bf D}_2)^T \vec{x} +  } \right. \nonumber \\
&&  + \left. \frac{i}{2\hbar} \vec{x}^T {  ({\bf C}_1{\bf B}_2 + {\bf D}_1 {\bf D}_2) ({\bf C}_1{\bf A}_2 + {\bf D}_1 {\bf C}_2)^T  } \vec{x} \right] .\label{GMutiplication}
\end{eqnarray}

We now plug in the expression for the ansatz given in (\ref{AnsatzKM}) and obtain that
\begin{eqnarray}
C^{(poly)}_{M_1 M_2 }(\vec{p}, \vec{p}') &=& k^{(poly)}_{M_1} k^{(poly)}_{M_2} \sqrt{ \frac{ \det({\bf D}_2 {\bf B}^{-1}_2)  }{\det({\bf D}_2 {\bf B}^{-1}_2 + { \bf D}^{-1}_1 {\bf C}_1 ) }  } e^{ -\frac{i}{2\hbar} \vec{p}^T ({\bf A}_1 {\bf B}_2 + {\bf B}_1 {\bf D}_2 ) ({\bf C}_1 {\bf B}_2 + {\bf D}_1 {\bf D}_2 )^{-1} \vec{p}} \sum_{ \vec{x}'} e^{ - \frac{i}{\hbar}\vec{x}'^T {\bf D}_2 \vec{p}'' } \times \nonumber \\
&& \times e^{ \frac{i}{\hbar}  \vec{x}'^T  {\bf D}_2 ({\bf C}_1 {\bf B}_2 + { \bf D}_1 {\bf D}_2 )^{-1} \vec{p}  } e^{ \frac{i}{2 \hbar} \vec{x}'^T \left[  {\bf D}_2 {\bf C}^T_2 + {\bf C}^T_1 {\bf D}^{-T}_1  - {\bf D}^{-1}_1 {\bf C}_1 ( {\bf D}_2 {\bf B}^{-1}_2 +  {\bf D}^{-1}_1 {\bf C}_1 ) {\bf D}^{-1}_1 {\bf C}_1     \right]\vec{x}' }.
\end{eqnarray}

By considering the change of variable $\vec{x}' = {\bf D}^{-T}_2 ( {\bf C}_1 {\bf B}_2 + {\bf D}_1 {\bf D}_2 )^T \vec{x}$ in the summation index $\vec{x}'$, the previous expression takes the form 
\begin{eqnarray}
C^{(poly)}_{M_1 M_2 }(\vec{p}, \vec{p}') &=& k^{(poly)}_{M_1} k^{(poly)}_{M_2} \sqrt{ \frac{ \det({\bf D}_2 {\bf B}^{-1}_2)  }{\det({\bf D}_2 {\bf B}^{-1}_2 + { \bf D}^{-1}_1 {\bf C}_1 ) }  } e^{ -\frac{i}{2\hbar} \vec{p}^T ({\bf A}_1 {\bf B}_2 + {\bf B}_1 {\bf D}_2 ) ({\bf C}_1 {\bf B}_2 + {\bf D}_1 {\bf D}_2 )^{-1} \vec{p}} \sum_{ \vec{x}} e^{ \frac{i}{\hbar}\vec{x}^T \vec{p} } \times \nonumber \\
&& \times e^{ -\frac{i}{\hbar}  \vec{x}^T  ({\bf C}_1 {\bf B}_2 + { \bf D}_1 {\bf D}_2 ) \vec{p}''  } e^{ \frac{i}{2 \hbar} \vec{x}^T (  {\bf C}_1 {\bf A}_2 + {\bf D}_1 {\bf C}_2)(  {\bf C}_1 {\bf B}_2 +  {\bf D}_1 {\bf D}_2  )^T \vec{x} }. \label{GMutiplication2}
\end{eqnarray}

We now equate both expressions (\ref{GMutiplication}) and (\ref{GMutiplication2}) and consider it also holds for the product ${\bf 1} = M \cdot M^{-1}$. This yields the following condition for the coefficient $k^{(poly)}_M$
\begin{equation}
k^{(poly)}_M =  \det ({\bf D} {\bf A}^T )^{-\frac{1}{4}}.
\end{equation}
This last coefficient is inserted on (\ref{AnsatzKM}) to give the final form for the kernel
\begin{equation}
C^{(poly)}_M(\vec{p}, \vec{p}') = \det ({\bf D} {\bf A}^T )^{-\frac{1}{4}} e^{ -\frac{i}{2 \hbar} \vec{p}^T  {\bf B} {\bf D}^{-1} \vec{p}} \sum_{\vec{x}} e^{ \frac{i}{\hbar} \vec{p}^T \vec{x} - \frac{i}{\hbar} \vec{p}'^T {\bf D}^T \vec{x} + \frac{i}{2\hbar} \vec{x}^T {  { \bf D} {\bf C}^T  } \vec{x}}, \label{KSpPoly}
\end{equation}
\noindent and this constitutes the main result of this section.

The first and most relevant aspect of the kernel (\ref{KSpPoly}) is that it gives rise to a non-unitary representation.  A unitary representation is achieved if the kernel in (\ref{KSpPoly}) satisfies 
\begin{equation}
\lim_{L} \frac{1}{2L} \int^L_{-L} d \vec{p} \; C^{(poly)}_M(\vec{p}, \vec{p}') C^{* (poly)}_M(\vec{p}, \vec{p}'') = \sum_{\vec{x}} e^{ - \frac{i}{\hbar} \vec{x}^T (\vec{p}' - \vec{p}'')}, \label{UnitarityCond}
\end{equation}
\noindent but this is only the case when $| k_M |^2 =1$ which implies that 
\begin{equation}
\det({\bf D} {\bf A}^T) = 1. \label{DetCond}
\end{equation}
\noindent This condition only holds for very special groups elements but not for the whole symplectic group, hence the representation is non-unitary.

The second aspect to be considered is the action of the subgroups mentioned before, $SO(n, \mathbb{R})$ and the scaling group. Using expression (\ref{KSpPoly}) for the rotation group we have 
\begin{eqnarray}
\widetilde{\Psi}(\vec{p})_{M(R)} &=& \lim_{L \rightarrow \infty} \frac{1}{2L}\int^L_{-L} d\vec{p}' C^{(poly)}_{M(R)}(\vec{p}, \vec{p}') \widetilde{\Psi}(\vec{p}')=  \lim_{L \rightarrow \infty} \frac{1}{2L}\int^L_{-L} d\vec{p}' \sum_{\vec{x}} e^{ \frac{i}{\hbar} \vec{p}^T \vec{x} - \frac{i}{\hbar} \vec{p}'^T {\bf R}^{-1} \vec{x}  } \widetilde{\Psi}(\vec{p}'), \nonumber \\
&=& \sum_{\vec{x}} e^{ \frac{i}{\hbar} \vec{p}^T \vec{x} } \Psi({\bf R}^{-1} \vec{x}) = \sum_{{\bf R} \vec{x}} e^{ \frac{i}{\hbar} \vec{p}^T {\bf R} \vec{x} } \Psi(\vec{x}) = \sum_{\vec{x}} e^{ \frac{i}{\hbar} \vec{p}^T {\bf R} \vec{x} } \Psi(\vec{x}) = \widetilde{\Psi}({\bf R}^T \vec{p}),
\end{eqnarray}
\noindent which coincides with the representation given by Chiou in \cite{chiou2007galileo}. As for the scalings, we have a similar result
\begin{eqnarray}
\widetilde{\Psi}(\vec{p})_{M(S)} &=& \lim_{L \rightarrow \infty} \frac{1}{2L}\int^L_{-L} d\vec{p}' C^{(poly)}_{M(S)}(\vec{p}, \vec{p}') \widetilde{\Psi}(\vec{p}')=  \lim_{L \rightarrow \infty} \frac{1}{2L}\int^L_{-L} d\vec{p}' \sum_{\vec{x}} e^{ \frac{i}{\hbar} \vec{p}^T \vec{x} - \frac{i}{\hbar} \vec{p}'^T {\bf S}^{-1} \vec{x}  } \widetilde{\Psi}(\vec{p}'), \nonumber \\
&=& \sum_{\vec{x}} e^{ \frac{i}{\hbar} \vec{p}^T \vec{x} } \Psi({\bf S}^{-1} \vec{x}) = \sum_{{\bf S} \vec{x}} e^{ \frac{i}{\hbar} \vec{p}^T {\bf S} \vec{x} } \Psi(\vec{x}) = \sum_{\vec{x}} e^{ \frac{i}{\hbar} \vec{p}^T {\bf S} \vec{x} } \Psi(\vec{x}) = \widetilde{\Psi}({\bf S} \vec{p}).
\end{eqnarray}
\noindent Both subgroups are unitarily represented. As we will see later, this is also the case for the polymer free particle but not for the harmonic oscillator.

\section{Propagators in polymer quantum mechanics} \label{S5}

In this section we determine the expressions for the propagators of the polymer free particle and the harmonic oscillator using (\ref{KSpPoly}). We showed that in the case of the harmonic oscillator, the propagator yields non-unitary evolution of any arbitrary polymer state (\ref{Pstate}).

\subsection{Free particle}

The propagator for the free particle in a regular lattice was given by Ernesto Flores et al. in \cite{flores2013propagators}. There, the propagator $G^{(EF)}(p,p')$  was derived using the eigenvalues solutions of the polymer free particle Hamiltonian. The propagator in momentum representation takes the form
\begin{equation}
G^{(EF)}(p,p') = \frac{2 \pi \hbar}{\mu} e^{- \frac{i}{\hbar} E_{p'} (t - t')} \delta(p-p'), \label{ErnestoPropagator}
\end{equation}
\noindent where the energy eigenvalue $E_p$ is of the form
\begin{equation}
E_p = \frac{\hbar^2}{m \mu^2} \left[ 1 - \cos\left( \frac{\mu p}{\hbar}\right)  \right].
\end{equation}

The propagator (\ref{ErnestoPropagator}) provides the time evolution of an arbitrary polymer state, say,
\begin{equation}
\widetilde{\Psi}^{(rl)}(p) = \sum_j \Psi_{n_j} e^{ i \frac{p n_j \mu}{\hbar}}, \label{RLstate}
\end{equation}
\noindent where $\{ n_j \}^{\infty}_{j=1}$ is a countable set of integers such that the set $\{ \mu n_j \}^{\infty}_{j=1}$ gives a graph with a countable number of elements. The graph constructed in this way is called {\it regular lattice graph} and the states defined on this graph are abreviately written as $\widetilde{\Psi}^{(rl)}(p)$. The polymer scale $\mu$ is, as usual, considered to be a small and arbitrary parameter. 

The time evolution of the state (\ref{RLstate}) using (\ref{ErnestoPropagator}) is 
\begin{eqnarray}
\widetilde{\Psi}^{(rl)}(p, t) &=& \int^{+ \frac{\pi \hbar}{\mu}}_{- \frac{\pi \hbar}{\mu}} G^{(EF)}(p,p') \widetilde{\Psi}(p') dp' = \int^{+ \frac{\pi \hbar}{\mu}}_{- \frac{\pi \hbar}{\mu} }e^{- \frac{i}{\hbar} E_{p'} (t - t')} \delta(p-p') \widetilde{\Psi}(p') dp', \nonumber \\
&=& e^{- \frac{i}{\hbar} E_{p} t }  \widetilde{\Psi}^{(rl)}(p). \label{ErnestoFPEvolved}
\end{eqnarray}

Let us analyze this system in our approach. First, recall the expression for ${\bf M}_{FP}(\Delta t)$ given in (\ref{PLP}), where $A=D=1$, $C=0$ and $B=\Delta t/m$ together with $n=1$. Plugging these values in (\ref{KSpPoly}) we obtain that the propagator for the free particle is given by
\begin{equation}
G^{(poly)}_{FP}(p, t; p' , t') = e^{-\frac{i}{2 \hbar} \frac{ (t - t') }{m} p^2 } \sum_x e^{ \frac{i}{\hbar} (p - p')x }, \label{FPFpropagator}
\end{equation}
\noindent which, when acting on a given arbitrary state in the full polymer Hilbert space of the form
\begin{equation}
\widetilde{\Psi}(p) = \sum_{ \{ x_j \}} \Psi_{ x_j} e^{ i \frac{p x_j }{\hbar}} \label{FState}
\end{equation}
\noindent gives
\begin{eqnarray}
\widetilde{\Psi}(p, t) &=& \lim_{L \rightarrow \infty} \frac{1}{2 L} \int^L_{-L} G^{(poly)}_{FP}(p, t; p' , t') \widetilde{\Psi}(p', t') dp' = e^{-\frac{i}{2 \hbar} \frac{ (t - t') }{m} p^2 } \sum_{ \{ x_j \} }  \Psi_{x_j} e^{ \frac{i}{\hbar} p x_j  } , \nonumber \\ 
 &=&e^{-\frac{i}{2 \hbar} \frac{ (t - t') }{m} p^2 } \widetilde{\Psi}(p) . \label{TEvolutionFP}
\end{eqnarray}
\noindent where we have taken $t' =0 $ for simplicity.

A global phase appears on both results (\ref{ErnestoFPEvolved}) and (\ref{TEvolutionFP}) and notice that the time-evolution is norm preserving due to ${\bf A} = {\bf D} = {\bf 1}$, i.e., condition (\ref{DetCond}) holds. Additionally, (\ref{ErnestoFPEvolved}) corresponds to a regular lattice, i.e., a polymer scale is considered whereas (\ref{TEvolutionFP}) corresponds to time evolution in the full polymer Hilbert space with no polymer scale. In this construction we used an unmodified Hamiltonian or energy spectrum, responsible of time evolution (\ref{TEvolutionFP}).

We can recover (\ref{ErnestoFPEvolved}) from expression (\ref{TEvolutionFP}). To do so, we have to consider a polymer scale $\mu$ which allows us to re-write an arbitrary graph $\{ x_j \} $ as the uncountable union of regular lattices centered in $\lambda$. That is to say, every point $x_j$ can be written as 
\begin{equation}
x_j = \lambda + \mu n_j, \label{decomposition}
\end{equation}
\noindent where $\lambda \in [0, \mu)$ and $n_j \in \mathbb{Z}$. As a result, this condition modifies the expression for (\ref{KSpPoly}) because it restricts the sum in (\ref{KSpPoly}).

Using these decomposition for the points $x_j$ of the state in (\ref{FState}) and fixing $\lambda=0$ gives 
\begin{equation}
\widetilde{\Psi}(p) = \sum_{ \{ j \}} \Psi_{ n_j} e^{ \frac{i }{\hbar} p \mu n_j}, \label{ScaleState1}
\end{equation}
\noindent which takes the same form as the expression for the state (\ref{RLstate}). By doing the same on the state (\ref{TEvolutionFP}) we have
\begin{equation}
\widetilde{\Psi}(p, t) = e^{-\frac{i}{2 \hbar} \frac{ (t - t') }{m} p^2 } \sum_{ \{ j \} }  \Psi_{n_j} e^{ \frac{i}{\hbar} \mu p n_j  },
\end{equation}
\noindent if we now consider the approximation for the square momentum term as
\begin{equation}
p^2 \rightarrow  \frac{2 \hbar^2}{\mu^2} \left[ 1 - \cos\left( \frac{\mu p}{\hbar}\right) \right], \label{Approximation}
\end{equation}
\noindent the resulting state coincides with (\ref{ErnestoFPEvolved}). In this way, the time-evolution provided by the polymer Hamiltonian considered for the free particle in \cite{flores2013propagators} is obtained.

This result has the following importance: the time evolution provided by the unitary action of the symplectic group in the full polymer Hilbert space (\ref{TEvolutionFP}) can be used to derive time evolution using a polymer scale (\ref{ErnestoFPEvolved}), that is to say, using the eigenvalues and eigenstates of the Hamiltonian  (\ref{SPProposal}). Naturally, to do this we followed an heuristic procedure which we summarize as follows:

\begin{enumerate} 
\item Calculate the propagator in the full polymer Hilbert space (\ref{FPFpropagator}).
\item Time-evolve an arbitrary polymer state (\ref{FState}) to obtain a new state (\ref{TEvolutionFP}).
\item Introduce the polymer scale $\mu$ and `decompose'  the graph points $\{ x_j \}$ of the state (\ref{FState}) as in (\ref{decomposition}). Instead of working on arbitrary lattices, consider only those centered in zero, i.e., $\lambda = 0$.
\item Replace the $p^2$ term in the global phase using the approximation in (\ref{Approximation}).
\end{enumerate}

This four steps are the ones we consider can led to derive the propagator of the polymer harmonic oscillator which is the goal of the next subsection.

\subsection{Quantum harmonic oscillator}

The first step leads to consider the symplectic matrix ${\bf M}_{HO}(t)$ in (\ref{SMHO}) where $A=D=\cos(\omega t)$ and $B=\frac{1}{m \omega} \sin(\omega t)$ and $C= - m \omega \sin(\omega t)$. With these values, the propagator in the full polymer Hilbert space is given as
\begin{equation}
G^{(poly)}_{HO}(p,t ; p',t') =  \frac{e^{ - \frac{i}{2 \hbar} \frac{ \tan(\omega t)}{m \omega}p^2 } }{   \sqrt{ \cos(\omega t)}  } \sum_x e^{ \frac{i}{\hbar} \left( p - p' \cos(\omega t) \right) x - \frac{i}{4 \hbar} m \omega \sin(\omega t) x^2  }. \label{PropFHO}
\end{equation}
\noindent We now proceed with the second step which requires the time evolution of an arbitrary state, say a polymer state of the form (\ref{FState}). The general expression is
\begin{eqnarray}
\widetilde{\Psi}(p,t) &=& \lim_{L \rightarrow \infty} \frac{1}{2L} \int^L_{-L} G^{(poly)}_{HO}(p,t ; p', t') \widetilde{\Psi}(p', t') dp',
\end{eqnarray}
\noindent and the time evolved state takes the following form
\begin{equation}
\widetilde{\Psi}(p,t) = \frac{e^{ - \frac{i}{2 \hbar} \frac{ \tan(\omega t)}{m \omega}p^2 } }{   \sqrt{ \cos(\omega t)}  } \sum_{ \{ x_j \}} \Psi_{ x_j} e^{ \frac{i}{\hbar}  \frac{x_j p }{\cos(\omega t)} } e^{  - \frac{i}{ 4 \hbar}  \frac{m \omega \sin(\omega t)}{\cos^2(\omega t)} x^2_j }. \label{TEHOFPolmyer}
\end{equation}

We now introduce the polymer scale $\mu$ and re-write the graph points using (\ref{decomposition}) and then take $\lambda =0$. Using this, the former state takes the form
\begin{equation}
\widetilde{\Psi}(p,t) = \frac{e^{ - \frac{i}{2 \hbar} \frac{ \tan(\omega t)}{m \omega}p^2 } }{   \sqrt{ \cos(\omega t)}  } \sum_{ \{ j \}} \Psi_{ n_j} e^{ \frac{i}{\hbar}  \frac{ \mu n_j p }{\cos(\omega t)} } e^{  - \frac{i}{ 4 \hbar}  \frac{m \omega \sin(\omega t)}{\cos^2(\omega t)} \mu^2 n^2_j }.
\end{equation}
Finally, as part of the fourth step, we approximate the square term $p^2$ as given in (\ref{Approximation}). The final expression for the time evolution of the polymer harmonic oscillator is
\begin{equation}
\widetilde{\Psi}(p,t) = \frac{e^{ \frac{ \tan(\omega t)}{i m \omega} \frac{ \hbar}{\mu^2} \left[ 1 - \cos\left( \frac{\mu p}{\hbar}\right) \right] } }{   \sqrt{ \cos(\omega t)}  } \sum_{ \{ j \}} \Psi_{ n_j} e^{ \frac{i}{\hbar}  \frac{ \mu n_j p }{\cos(\omega t)} } e^{  - \frac{i}{ 4 \hbar}  \frac{m \omega \sin(\omega t)}{\cos^2(\omega t)} \mu^2 n^2_j }. \label{TEHORLPolymer}
\end{equation}
This result together with expressions (\ref{TEHOFPolmyer}) or (\ref{PropFHO}) are the main results of this subsection, and are also the relevant results of this work. As can be seen, the time-evolution in both scenarios (\ref{TEHOFPolmyer}) and (\ref{TEHORLPolymer}) is not unitary because condition (\ref{DetCond}) does not hold.

\section{Discussion} \label{S6}

In this paper, we presented the representation of the symplectic group in polymer quantum mechanics (\ref{KSpPoly}). We paid particular attention to those linear transformations generating time evolution of linear systems such as the free particle and the harmonic oscillator and in this way, we derived their quantum propagators. However, the representation is not unitary although, a subset of the symplectic group with elements such that $\det {\bf A} \det {\bf D} =1$ admit a unitary representation. Among the subgroup unitarily represented we found the rotation group $SO(n, \mathbb{R})$ and the scaling group formed by non-singular diagonal matrices. 

The propagator for the free particle, was derived in the full polymer Hilbert space (\ref{FPFpropagator}) and the time-evolution it provides on an arbitrary state was also calculated in (\ref{TEvolutionFP}). We re-derived the result in \cite{flores2013propagators} for the polymer free particle by implementing some procedures within the full polymer Hilbert space to obtain the regular lattice description. This procedure served later to obtain the time-evolution for the polymer harmonic oscillator. On the other hand, time-evolution for the polymer free particle on both, the full polymer Hilbert space and the regular lattice description is unitary. Remarkably, time-evolution for this system is consistent with the Hamiltonian in (\ref{SPProposal}) instead of Hamiltonian (\ref{EPProposal}). Further analysis is required to explore why this representation selects one polymer Hamiltonian over the other options.

We determined the propagator for the polymer harmonic oscillator in (\ref{PropFHO}) and then we used it to obtain the time-evolution of an arbitrary state in the full polymer Hilbert space (\ref{TEHOFPolmyer}). Finally, we applied the procedures derived in the polymer free particle case to obtain the time-evolution in a regular lattice (\ref{TEHORLPolymer}). Due to condition (\ref{DetCond}) does not hold the time-evolution is not unitary in the case of the polymer harmonic oscillator. This contradicts the unitary time-evolution using the exponentiation of the polymer Hamiltonian in (\ref{SPProposal}). Moreover, the time-evolved state in (\ref{TEHORLPolymer}) seems to violate the superselection rules due to the time-dependency in the phase $e^{ \frac{i}{\hbar}  \frac{ \mu n_j p }{\cos(\omega t)} }$ although more analysis to confirm this is needed.

Finally, these results might suggest that in any limiting procedure to describe or recover the standard quantum mechanics for these two systems, the free particle and the harmonic oscillator, should also recover the unitary representation of the symplectic group. 

\section{Acknowledgments}
I thank H. Morales-T\'ecotl, and Viqar Husain for useful discussions, and to Martin Bojowald for his comments on a preliminary version of this work.


\end{document}